\documentclass[a4paper,fleqn,usenatbib]{mnras}

\usepackage{newtxtext,newtxmath}
\usepackage[T1]{fontenc}
\usepackage{ae,aecompl}
\usepackage{hyperref}
\usepackage[dvipdfmx]{graphicx}	
\usepackage{amsmath}	
\usepackage{amssymb}	
\usepackage{bm}

\title[Detectability of Galactic Isolated Black Holes]{X-ray Detectability of Accreting Isolated Black Holes in Our Galaxy}
\author[D.Tsuna et al.]{
Daichi Tsuna,$^{1,2}$\thanks{tsuna@resceu.s.u-tokyo.ac.jp}
Norita Kawanaka,$^{3,4}$
and Tomonori Totani$^{5,1}$\\
$^{1}$Research Center for the Early Universe (RESCEU), the University of Tokyo, Hongo, Tokyo 113-0033, Japan\\
$^{2}$Department of Physics, School of Science, the University of Tokyo, Hongo, Tokyo 113-0033, Japan\\
$^{3}$Department of Astronomy, Graduate School of Science, Kyoto University, Kitashirakawa Oiwake-cho, Sakyo-ku Kyoto, 606-8502, Japan\\
$^{4}$Hakubi Center, Yoshida Honmachi, Sakyo-ku, Kyoto 606-8501, Japan
\\
$^{5}$Department of Astronomy, School of Science, the University of Tokyo, Hongo, Tokyo 113-0033, Japan}
\date{Accepted XXX. Received YYY; in original form ZZZ}
\pubyear{2018}

\begin{document}
\label{firstpage}
\pagerange{\pageref{firstpage}--\pageref{lastpage}}
\maketitle

\begin{abstract}
Detectability of isolated black holes (IBHs) without a companion star
but emitting X-rays by accretion from dense interstellar medium (ISM)
or molecular cloud gas is investigated.  We calculate orbits of IBHs
in the Galaxy to derive a realistic spatial distribution of IBHs, for
various mean values of kick velocity at their birth $\upsilon_{\rm
  avg}$.  X-ray luminosities of these IBHs are then calculated
considering various phases of ISM and molecular clouds, for a wide
range of the accretion efficiency $\lambda$ (a ratio of the actual
accretion rate to the Bondi rate) that is rather uncertain. It is
found that detectable IBHs mostly reside near the Galactic Centre
(GC), and hence taking the Galactic structure into account is
essential. In the hard X-ray band, where identification of IBHs from
other contaminating X-ray sources may be easier, the expected number
of IBHs detectable by the past survey by {\it NuSTAR} towards GC is at
most order unity. However, 30--100 IBHs may be detected by the future
survey by {\it FORCE} with an optimistic parameter set of
$\upsilon_{\rm avg} = 50 \ \mathrm{km\ s^{-1}}$ and $\lambda = 0.1$,
implying that it may be possible to detect IBHs or constrain the model parameters.
\end{abstract}


\section{Introduction}

A black hole is thought to form in the last stage of stellar
evolution, when a massive star gravitationally collapses at the end of
its life. Owing to the development of X-ray observation technology,
more than 20 strong black hole candidates have been detected in our
Galaxy as X-ray binaries (see, e.g.  \citealt{RM06} for a review). An
even stronger proof of the existence of stellar mass black holes has
been obtained by the recent detections of gravitational waves from binary mergers of two black holes \citep{GW150914,GW151226,GW170104,GW170814,GW170608}.
However it is expected that there are many more black holes without a
companion star, which are often called isolated black holes (IBHs).
The number of black holes that formed in the Milky Way in the past is
estimated to be $\sim 10^8$, based on stellar evolution theory, total
stellar mass of the Galaxy, observed metallicity and chemical
evolution modeling \citep[e.g.,
][]{Shapiro83,vandenHeuvel92,Samland98,Caputo17}. 

IBHs, which are probably occupying more than half of the total black
hole population \citep{Fender13}, are expected to shine by accreting
surrounding gas. If mass accretion onto IBHs can be described by the
Bondi spherical accretion formula \citep{HL39,BH44,Bondi52}, the
accretion rate is proportional to $\rho \upsilon^{-3}$, where $\rho$
and $\upsilon$ are density of surrounding gas and IBH velocity,
respectively. Therefore IBHs that plunge into a dense gas cloud (which
mainly exists in the Galactic disc) with a sufficiently low velocity
can obtain a large mass accretion.  Such IBHs may be brighter than
isolated neutron stars (INSs), because of heavier masses and slower
velocities, though INSs are expected to be more abundant than IBHs by
an order of magnitude \citep{Arnett89,ST10}.

There have been many studies on the detectability of such accreting
INSs \citep{Ostriker70,Treves91,BM93,Treves00,Perna03,Ikhsanov07} and
IBHs
\citep{Shvartsman71,Grindlay78,Carr79,McDowell85,CP93,PP98,Fujita98,Armitage99,Grindlay01,Agol02,Maccarone05,MT05,Sartore10,Motch12,Barkov12,Fender13,Ioka17,Matsumoto17}. 
A number of observational
searches have also been performed in the past for such accreting INSs
and IBHs \citep{Stocke95,Walter96,Wang97,Schwope99,Chisholm03,Muno06}.
Although several INSs candidates have been identified by their thermal
emission (e.g. ``The Magnificent Seven'' identified by the {\it ROSAT}
satellite; see e.g. \citealp{Haberl07,Kaplan08} for reviews), no
accretion-powered INSs or IBHs have been detected so far.

In estimates of accretion-powered IBH detectability, there are some
sources of uncertainty.  One is the natal kick velocity $\upsilon_{\rm
  kick}$ of black holes. 
In spite of intense theoretical and observational studies
(\citealt{WvP,JN04,Gualandris05,Miller-Jones09,Fryer12,Repetto12,Reid14,Wong14,Repetto15,Mandel16,Wysocki17}; for a recent extensive review see \citealt{Belczynski16}), the
distribution and mean value of $\upsilon_{\rm kick}$ (hereafter denoted as $\upsilon_{\rm avg}$) are still highly
uncertain. BH kick velocities may be smaller or comparable to that of
neutron star kicks (typically a few hundred km\ s$^{-1}$; \citealt{Hobbs05}) that can be inferred from proper motion of
pulsars. Another source of uncertainty is the efficiency of accretion,
$\lambda$, which is the ratio of actual accretion rate onto IBHs to
the Bondi rate. The spherical Bondi accretion formula can be applied
only when angular momentum of accretion flow is negligible.  Once
angular momentum becomes important, the accretion flow should be
described by other modes, such as radiatively inefficient accretion
flow (RIAF) \citep{Ichimaru77,NY95,KFM08}. It is generally expected that accreting matter is lost by outflow on various scales, and hence the actual accretion reaching to BH horizons can be significantly reduced.  In other words,
observational searches for IBHs will give constraints on these
parameters of $\upsilon_{\rm avg}$ and $\lambda$.

The aim of this paper is to make the best estimate currently possible
for the number and luminosities of IBHs that are formed by normal
stellar evolution and shining by accretion from ISM.  We then evaluate
detectability by the past and present X-ray observations such as {\it ROSAT} and {\it NuSTAR}, and
especially we focus on the prospect of a future project {\it FORCE}
\citep{FORCE} in hard X-ray band.  Since luminous IBHs are expected to
be found in dense gas regions, hard X-ray observation would be
particularly powerful because of less absorption. Discrimination of
IBHs from other populations of X-ray sources is also important, and
the hard X-ray band is useful to examine the spectral difference
between IBHs and cataclysmic variables \citep{RM06,Nobukawa16}. In order to obtain a more reliable estimate
than previous studies, we calculate the spatial distribution of IBHs by
solving their orbits in the gravitational potential of the Galaxy with
various values of kick velocities.  Then accretion rate is estimated
by using a realistic gas distribution in the Galaxy based on latest
observations.

The paper is organized as follows. In Section \ref{sec:formulations}
we give formulations of our calculations, with discussions about the plausible ranges of $\upsilon_{\rm avg}$ and $\lambda$. Then main results, including the expected number of detectable IBHs for the wide ranges of $\upsilon_{\rm avg}$ and $\lambda$, are presented in Section \ref{sec:results} with prospects of future observations. After some discussion on the caveats in our work in Section \ref{sec:discussion} we conclude in Section \ref{sec:conclusions}.


\section{Formulations}
\label{sec:formulations}


\subsection{The Galactic Structure}

In a realistic picture within the framework of cosmological galaxy and
structure formation driven by cold dark matter, the dynamical
structure of our Galaxy should have evolved in time
(e.g. \citealt{Amores17}).  However, it is difficult to construct a
reliable model of accurate time evolution for our Galaxy, and here we
use a simplified model in which the Galactic structure does not
evolve, and is composed of three components: the central bulge, the
Galactic disc, and a spherical dark halo surrounding them.  First we
introduce the model of the Galactic gravitational potential that we
use, and then we present our model of IBH birth location and
history.  Here we use a cylindrical coordinate system of radius,
azimuth, and height ($r,\theta,z$), with the $z$-direction
perpendicular to the Galactic plane.

We follow the gravitational potential model of \citet{Irrgang13} (the
Model II).  For the gravitational potential of the spherical bulge and
disc, the model proposed by \citet{MN75} is assumed, which is described as
\begin{align}
\phi_i(r,z)=-\frac{GM_i}{\sqrt{r^2+\left(a_i+\sqrt{z^2+b_i^2}\right)^2}}.
\label{eq:potential_bd}
\end{align}
where $i=1,2$ represent the bulge and disc respectively.  The
spherical model of \citet{Wilkinson99} is assumed for the dark halo
potential:
\begin{align}
\phi_{\mathrm{halo}}=-\frac{GM_h}{R_h}\ln\left(\frac{\sqrt{R^2+R_h^2}+R_h}{R}\right),
\label{eq:potential_h}
\end{align}
where $R\equiv \sqrt{r^2+z^2}$. \citet{Irrgang13} obtained the values
of the constants $M_i,a_i,b_i,M_h,$ and $R_h$ by comparison with
observational data, e.g. the Galactic rotation curve and the mass
densities in the solar vicinity. These are:
\begin{align}
M_1&=4.07\times 10^{9}\ \mathrm{M_{\sun}},\ a_1=0,\ b_1=0.184\ \mathrm{kpc} \\
M_2&=6.58\times 10^{10}\ \mathrm{M_{\sun}},\ a_2=4.85\ \mathrm{kpc},\ b_2=0.305\ \mathrm{kpc}\\
M_h&=1.62\times 10^{12}\ \mathrm{M_{\sun}},\  R_h=200\ \mathrm{kpc}.
\label{eq:value_Galaxy}
\end{align}

We assume that the IBH birth rate per unit volume has an exponential
radial profile, $\rho_{\rm IBHb} \propto \exp( - r / r_d )$ with the
scale length $r_d = $ 2.15 kpc of the stellar disc \citep{Licquia15}.
Along the disc height $\rho_{\rm IBHb}$ is assumed to be uniform in
the region of $|z| < h$, where we adopt $h = 75$ pc which is the scale
height of molecular clouds in the Galaxy (see Table \ref{tab:ISMtable}
in Section \ref{ssec:ISM}).  A spherical exponential profile is
assumed for $\rho_{\rm IBHb}$ in the bulge, as $\rho_{\rm IBHb}
\propto \exp(-R/R_b)$ and $R_b = $ 120 pc, following \citep{Sofue13},
who found that the spherical exponential profile fits
better to the observed data than the de Vaucouleurs law that is
conventional as the profile of spheroidal galaxies.

The total number of IBHs born in the past in the Galaxy is set to
$N_{\rm IBH} = 1 \times 10^8$, and of course the final number of the
observed IBHs found in this work scales with this parameter.  Recent
observations show that the Galactic bulge contains $\sim 15$ per cent
of the total stellar mass in the Galaxy, while the remaining $\sim 85$
per cent are in the disc \citep{Licquia15}. Hence we determine the IBH
birth rate in the disc and bulge so that the disc-to-bulge number
ratio of the total IBH numbers is the same as that of stellar mass.
This is a reasonable assumption provided that the initial mass
function (IMF) does not depend on time or location throughout the
Milky Way (for a discussion on IMF variability over cosmic history and
within the Milky Way, see \citealt{Bastian10} and \citealt{Wegg17}
respectively). The IBH birth rate in the disc is assumed to be
constant with time from 10 Gyrs ago to now.  On the other hand, the
bulge is composed mostly by old stellar populations, which were
presumably formed in the early stage of the history of the Galaxy.
Quantitative star formation history of the bulge is still under debate
(see \citealt{Nataf16} for a review), and here we assumed a simple
history that bulge stars formed uniformly in a time period of $2$ Gyrs
spanning from $10$ to $8$ Gyrs ago, which is within the range of
uncertainty about the bulge star formation history.


\subsection{IBH Initial Conditions}

Here we describe the initial conditions of IBHs, namely the BH mass
distribution and initial velocities.  We adopt the IBH mass
distribution of \citet{Ozel10} obtained from observation of X-ray
binaries: a normal distribution of average $7.8\ \mathrm{M_{\sun}}$
and standard deviation $1.2\ \mathrm{M_{\sun}}$. This distribution is
consistent with a parallel Bayesian estimation by \citet{Farr11}.  It
is assumed that the mass change due to accretion is negligible.

The initial velocity of an IBH is calculated as the sum of the
velocity of the progenitor star and the kick velocity given at the
time of the IBH formation. Velocities of IBH progenitors formed in the
disc are assumed to follow the rotation velocity of the Milky Way
consistent with the potential model of \citet{Irrgang13}, which is approximated as
\begin{align}
\upsilon_\phi=
\begin{cases}
265-1875(r-0.2)^2&\mathrm{km\ s^{-1}}\ \ \ ($for$\  r<0.2)\\
225+15.625(r-1.8)^2&\mathrm{km\ s^{-1}}\ \ \ ($for$\  0.2<r<1.8)\\
225+3.75(r-1.8)&\mathrm{km\ s^{-1}}\ \ \ ($for$\  1.8<r<5.8)\\
240&\mathrm{km\ s^{-1}} \ \ \ ($for$\  r>5.8)
\end{cases}
\label{eq:rotation_curve}
\end{align}
where $r$ is measured in kpc.  The motion of stars in the bulge is
instead dominated by random motion rather than rotation.  Thus we
assume that the progenitors of IBHs in the bulge have a Maxwell-Boltzmann
velocity distribution with the mean of $130\ \mathrm{km\ s^{-1}}$,
which is consistent with velocity measurements of stars located near
the Galactic Centre (hereafter GC) \citep{Kunder12}.

The kick velocity distribution of BHs is hardly known.  It is often
supposed that a natal kick speed decreases with BH mass, as expected
in the case of a fixed momentum. This implies a reduced BH kick speed
compared to neutron stars \citep[e.g.,][]{Fryer12}. A study of 233 pulsars by \citet{Hobbs05} concluded that neutron star kicks
obey a Maxwell-Boltzmann distribution with 1D standard deviation
$\sigma = 265\ \mathrm{km\ s^{-1}}$, which corresponds to an average
3D kick velocity of about $420\ \mathrm{km\ s^{-1}}$. If we simply
extrapolate this result to black holes with conserved momentum, we
obtain an average kick speed as low as $50\ \mathrm{km\ s^{-1}}$.
However some studies \citep{Repetto12,Repetto15} claim that this may
not be the case, and the present locations of some X-ray binary
systems require a natal kick broadly in the range of 100--500
km\ s$^{-1}$, which is comparable to neutron star kicks.
\citet{Janka13} proposed a possible theoretical explanation about
this result. This is still a matter of debate, and here we assume a
Maxwell-Boltzmann kick velocity distribution with a 3D average
velocity $\upsilon_{\mathrm{avg}}$ in the range of 50--400 km/s,
and we assume that the kick speed is not correlated with 
BH mass for simplicity.

If the initial kick velocity is generated by the Monte-Carlo method
obeying the Maxwell-Boltzmann distribution, the probability of getting
a velocity much lower than the average is small, but such IBHs have a
high chance of detection by higher Bondi accretion rate.  Because of
the limitation of computing time, the number of orbital calculations
($N_{\rm MC} \sim 10^6$) is much smaller than the actual IBH number
$N_{\rm IBH} \sim 10^8$, and the small velocity IBHs are not well
sampled by the Monte-Carlo generation. Therefore we set the grids of
kick velocity for the orbital calculation that is more uniform than
the Maxwell-Boltzmann distribution, and multiply the probability
distribution of the kick velocity in the final output of the
detectable number of IBHs (e.g., X-ray source counts).


\subsection{Equation of Motion}
\label{ssec:EQM}

IBHs formed and kicked in our Galaxy will move following the
Galactic gravitational potential. By simple calculation using
eq. \ref{eq:potential_bd} and eq. \ref{eq:potential_h} we obtain the
equations of motion as follows:
\begin{align}
\frac{dr}{dt}=&\upsilon_r\\
\frac{dz}{dt}=&\upsilon_z\\
\frac{d\upsilon_r}{dt}=&-\frac{\partial \Phi}{\partial r}+\frac{j_z^2}{r^3} \nonumber\\
=&\frac{j_z^2}{r^3}-\sum_{i=1,2}\frac{GM_ir}{\left\{r^2+[a_i+(z^2+b_i^2)^{1/2}]^2\right\}^{3/2}}\nonumber \\
&-\frac{GM_h}{R_h}\cdot\frac{r}{\sqrt{r^2+z^2}}\left[\frac{R_h}{\sqrt{r^2+z^2}\sqrt{r^2+z^2+R_h^2}} \right]\\
\frac{d\upsilon_z}{dt}=&-\frac{\partial\Phi}{\partial z} \nonumber \\
=&-\sum_{i=1,2}\frac{GM_iz[a_i+(z^2+b_i^2)^{1/2}]}{\left\{r^2+[a_i+(z^2+b_i^2)^{1/2}]^2\right\}^{3/2}\sqrt{z^2+b_i^2}}\nonumber \\
&-\frac{GM_h}{R_h}\cdot\frac{z}{\sqrt{r^2+z^2}}\left[\frac{R_h}{\sqrt{r^2+z^2}\sqrt{r^2+z^2+R_h^2}}\right].
\label{eq:eq_of_motion}
\end{align}
We have used $\Phi\equiv \phi_1+\phi_2+\phi_{\mathrm{halo}}$ to denote
the total Milky Way potential. Here $j_z\equiv r\upsilon_\theta$ is
the z-axis specific angular momentum, and since the potential $\Phi$ is
independent of $\theta$, $j_z$ will be conserved. 
Thus the rotational velocity $\upsilon_\theta$ can be obtained from the conservation of
$j_z$, which greatly simplifies our calculation. The four equations
are integrated using the 4th-order Runge-Kutta method, and as a result
the present location and velocity of each IBH are obtained.

Dynamical friction by stars or gas in molecular clouds
\citep{Ostriker99,MT05,Inoue17} is not considered in our calculation
of IBH orbits. These effects are larger for more massive black holes,
but negligible for stellar mass black holes.


\subsection{Interstellar Gas}
\label{ssec:ISM}

Once the present location and velocity of the IBHs are calculated, the
next information needed for estimating the accretion rate is the profile of
interstellar gas clouds in the Milky Way.  We consider five ISM phases
that differ by temperature and density \citep{BH00}. The densest are
the molecular clouds composed mostly of $\mathrm{H_2}$, followed by
the cold neutral medium mostly made up of cold H gas. These two types
of gases are expected to be the regions where accretion becomes large
enough to make IBHs observable. However the other three phases, the
warm neutral medium (warm H\,{\sevensize I}), warm ionized medium
(warm H\,{\sevensize II}), and hot ionized medium (hot H\,{\sevensize
  II}) occupy the majority of the volume.

Table \ref{tab:ISMtable} lists the ISM parameters around the solar
neighbourhood adopted in this work. For the densest two phases
(molecular clouds and cold H\,{\sevensize I}) we assume that the probability
distribution of gas particle density at a given point is described by
a power law with an index $\beta$
in the range $n_1 < n < n_2$ \citep{Agol02}.  Thus the
volume filling fraction of gas with density $n$ to $n + dn$
is expressed as $(d\xi / dn) dn$, and  
\begin{align}
\frac{d\xi (n)}{dn}=\frac{\beta-1}{n_1^{1-\beta}-n_2^{1-\beta}}
\, \tilde{\xi}(r, z) n\, ^{-\beta} \ \ (n_1<n<n_2) \ ,
\label{eq:fillingfactor_vs_density}
\end{align}
where $\tilde{\xi}(r, z)$ is the volume filling fraction integrated over
$n$. In calculation of final IBH numbers, we set many bins in $n_1 < n
< n_2$, and when an IBH is found in these ISM phases after the orbital
calculation, X-ray luminosity is calculated for each $n$ bin. Then
IBHs of the luminosity corresponding to varoius bins are summed up
with the probability distribution $d\xi/dn$ multiplied.

On the other hand, the other three phases (warm H\,{\sevensize I},
warm and hot H\,{\sevensize II}) are represented by a single gas
density from \citet{BH00}. The mid-plane volume filling fraction
around the solar neighbourhood \citep{BH00} and the disc scale heights
\citep{Agol02} of these five phases are shown as $\tilde{\xi}_0^{\rm BR}$ and
$H_d$, respectively, where the subscript 0 indicates the value at the solar neighbourhood.
Effective sound velocity, which includes turbulent velocity that
becomes dominant in cold phases, is also necessary to estimate the
Bondi accretion rate. This is set to $c_s=3.7(n/100\ {\rm
  cm}^{-3})^{-0.35}$ km\ s$^{-1}$ for molecular clouds \citep{MT05}
from the observational results of turbulent velocity by
\citet{Larson81}, $c_s=150$ km\ s$^{-1}$ for the hot H\,{\sevensize
  II} phase, and $10$ km\ s$^{-1}$ for the other three phases
\citep{Ioka17}.

It is assumed that the gas densities of each ISM phase are constant,
but volume filling fractions depend on the location in the Galaxy.
The filling fractions of molecular clouds, cold and warm
H\,{\sevensize I} phases are determined to match the surface density
profile $\Sigma(r)$ of $\mathrm{H_2}$ and H\,{\sevensize I} recently
obtained by \citet{NS16}.  We assume that the gas distribution along
the height from the disc plane is uniform in the region of $|z| <
H_d$, and $H_d$ is constant in the Galaxy.  Hence the filling fraction
can be calculated from the surface density as
\begin{align}
\tilde{\xi}(r, z) = \frac{\Sigma(r)}{2H_d\mu} \frac{1}{n} \ 
\label{eq:fillingfactor}
\end{align}
when $|z| < H_d$ but zero otherwise, where $\mu=2.72m_p$ and $1.36m_p$
for molecular and atomic gas clouds respectively, and $m_p$ is the
proton mass.  The gas density $n$ of warm H\,{\sevensize I} is simply
given in Table \ref{tab:ISMtable}, but that for molecular clouds and
cold H\,{\sevensize I} should be replaced by the mean density of the
power-law distribution,
\begin{align}
\langle n \rangle = \frac{1}{\tilde{\xi}}
\int_{n_1}^{n_2} n \frac{\partial \xi}{\partial n} dn 
= \frac{\beta-1}{\beta-2}
\frac{n_1^{2-\beta}-n_2^{2-\beta}}{n_1^{1-\beta}-n_2^{1-\beta}} \ .
\label{eq:fillingfactor}
\end{align}
Observed $\Sigma(r)$ of H\,{\sevensize I} includes both cold and warm
H\,{\sevensize I}, and we assume that the relative proportion of these
two phases is 3.1:3.5 and constant throughout the Galaxy, which is
calculated by $\Sigma_0 = 2 H_d \mu \tilde{\xi}_0^{\rm BR} n$ at the solar
neighbourhood with the parameters given in Table \ref{tab:ISMtable}.
Then $\tilde{\xi}(r, z)$ for the three phases of $\mathrm{H_2}$ and
H\,{\sevensize I} has been determined throughout the Galaxy.  It
should be noted that the filling fraction may become larger than the
unity depending on $\Sigma(r)$ in this formulation, but we confirmed
that the total of $\tilde{\xi}$ for these three phases is less than one
everywhere in the Galaxy with the observed values of $\Sigma(r)$.

Then the filling fractions $\tilde{\xi}(r, z)$ of the other two phases (warm
and hot H\,{\sevensize II}) are determined as follows.  At a given
location, the total filling fraction $\tilde{\xi}_{\rm H2+HI}$ of the three
phases ($\mathrm{H_2}$ and cold/warm H\,{\sevensize I}) is
calculated. (Note that some of the three phases do not exist depending
on the height $z$.)  Then the remaining filling fraction, $1 -
\tilde{\xi}_{\rm H2+HI}$, is distributed into two H\,{\sevensize II} phases
when $|z| \le 1$ kpc, assuming that the ratio of $\tilde{\xi}_0^{\rm BR}$ for
the solar neighbourhood is constant throughout the Galaxy.  In regions
of $1 < |z| \le 3$ kpc the fraction $1 - \tilde{\xi}_{\rm H2+HI}$ is filled
only by hot H\,{\sevensize II}.  

The filling fractions thus determined
in this work are different from $\tilde{\xi}_0^{\rm BR}$ even at a mid-plane
point of the Sun's Galactocentric distance ($r = R_0 = $ 8.3 kpc,
\citealt{Gillessen09,Russeil17}; this is consistent with
\citealt{Irrgang13} model II as well), which are also shown in Table
\ref{tab:ISMtable}.

\begin{table*}
\centering
\begin{tabular}{ccccccccc}
\hline 
Phase & $n_1\mathrm{[cm^{-3}]}$ & $n_2\mathrm{[cm^{-3}]}$ & $\beta$ 
& $\tilde{\xi}_0^{\rm BR}$ & $H_d$ & $c_s[\mathrm{km\ s^{-1}}]$ 
& $\Sigma_0 [\mathrm{M_{\sun}\ pc^{-2}}]$ 
& $\tilde{\xi}(r=8.3{\rm \ kpc})$ \\ \hline \hline
Molecular clouds & $10^2$ & $10^5$ & $2.8$ & 0.001 & $75$ pc & $3.7(n/100\ {\rm cm}^{-3})^{-0.35}$ & $2.3$ & $0.0004$\\
Cold H\,{\sevensize I} & $10^1$ & $10^2$ & $3.8$ & 0.02 & $150$ pc & $10$ & $3.1$ & $0.026$ \\
Warm H\,{\sevensize I} & \multicolumn{2}{|c|}{$0.3$} & -- & 0.35 & $500$ pc & $10$ & $3.5$ & $0.46$  \\
Warm H\,{\sevensize II} & \multicolumn{2}{|c|}{$0.15$} & -- & 0.20 & $1$ kpc & $10$ & $2.0$ & $0.16$ \\
Hot H\,{\sevensize II} & \multicolumn{2}{|c|}{$0.002$} & -- & 0.43 & $3$ kpc & $150$ & $0.17$ & $0.37 $ \\ \hline
\end{tabular}
\caption{The five ISM phases considered in this work.  A power-law
  probability distribution is assumed for molecular clouds and cold H\,{\sevensize I}
  in the range $n_1 < n < n_2$ with an index $\beta$, but a single
  density is assumed for the other three phases.  The mid-plane volume
  filling fractions around the solar neighbourhood \citep{BH00}, disc
  scale heights \citep{Agol02}, and effective sound velocities
  \citep{MT05,Ioka17} are shown as $\tilde{\xi}_0^{\rm BR}$, $H_d$, and $c_s$,
  respectively.  The mass surface density around the solar
  neighbourhood, $\Sigma_0$, is calculated from gas density,
  $\tilde{\xi}_0^{\rm BR}$, and $H_d$. The mid-plane values of the
filling fraction $\tilde{\xi}$ at the Sun's location
($r = $ 8.3 kpc) in our modelling are also shown.}
 \label{tab:ISMtable}
\end{table*}


\subsection{Mass Accretion from ISM}
There are studies that claim the accretion onto compact objects could
be much less than the Bondi accretion rate, due to material outflow in
the process of accretion. These claims are supported by theoretical
modelings of accretion flows including outflow \citep{BB99} and
hydrodynamical and MHD simulations (see \citealt{Perna03}).

These studies suggest that the ratio of the actual accretion to the
Bondi accretion, $\lambda$, scales as
$(R_{\mathrm{in}}/R_\mathrm{out})^p$, where $R_{\mathrm{in}}$ and
$R_\mathrm{out}$ are the inner and outer radius of the accretion
flow respectively, and index $p$ being an uncertain number
around $0.5-1$ \citep{Yuan14}. The inner radius $R_{\mathrm{in}}$ is
generally expected to be about a few to few tens times the
Schwartzschild radius, but $R_{\mathrm{out}}$ should be dependent on 
the angular momentum at the Bondi radius. 

If angular momentum is sufficiently large to make the accretion flow
rotationally supported at the Bondi
radius, $R_{\mathrm{out}}$ will be comparable to the Bondi
radius. This is for the case of Sgr A*, as assumed by several authors
\citep{Yuan03,Totani06}, and in this case
$R_{\mathrm{in}}/R_\mathrm{out}$ will be extremely small, down to
$10^{-8}-10^{-9}$. If we simply adopt this and use $p=0.5-1$, we get
$\lambda$ no higher than $10^{-4}$. However for Sgr A* a smaller index
$p=0.27$ is preferred from fit to observations \citep{Yuan03}, which
gives $\lambda \sim 0.01$. This agrees with the observation of nearby
active galaxies by \citet{Pellegrini05}, who estimated 
$\lambda$ to be around $0.01$. We do not know, however, whether this
can be applicable to stellar-mass black holes, since their Bondi radii
are very different in scale from supermassive black holes.
 
The study by \citet{Perna03} discusses the case for accretion onto
INSs, which concluded that $\lambda \lesssim 10^{-3}$ is
consistent with the null detection of accreting INSs by {\it
  ROSAT}. However, neutron stars have magnetic fields and hard
surfaces which significantly affect the accretion rate
\citep[e.g.,][]{Toropina12}. Thus we cannot simply assume that IBHs would follow
this constraint.

Some studies
\citep[e.g.,][]{Fujita98,Agol02,Ioka17,Matsumoto17,Inoue17} used
observations of the density \citep{Armstrong95} or velocity
fluctuations \citep{Larson81} of the interstellar medium, deriving
that $R_{\mathrm{out}}$ is much smaller than the Bondi radius. They
obtain $R_{\mathrm{out}}\sim 10^5 R_{s}$, which gives a range
$\lambda=10^{-4}$--$10^{-2}$ for $p=0.5$--$1$. However the
observational results for ISM density fluctuations include significant
uncertainties, and, more importantly, the observation by
\citet{Armstrong95} targets ionized hot gas. Thus we cannot apply this
relation in the case when IBHs accrete neutral molecular gas,
which is the most observable case. Estimates based only on ISM velocity
fluctuations would be a lower limit of the initial angular momentum,
because it would increase by density fluctuations and the velocity
of the black hole. The IBH velocity is typically much larger than the
turbulent velocity of the interstellar medium.

To summarize, there is a large uncertainty about the
accretion efficiency $\lambda$ both theoretically and
observationally. Therefore here we simply test $\lambda$ in the range
of $10^{-3}$--$10^{-1}$. Although this may be rather optimistic, we
adopt this because later we will find in our calculations that IBHs may be detectable by future surveys only when $\lambda\gtrsim 0.01$.

\subsection{IBH Luminosity and Flux}
\label{ssec:lum_flux}

We estimate the bolometric luminosity $L$ from the Bondi-Hoyle
accretion rate with the $\lambda$-factor, as
\begin{align}
\dot{M}=&\lambda \cdot
4\pi\frac{(GM)^2\rho}{(\upsilon^2+c_s^2)^{3/2}}\nonumber \\ \approx&
3.7\times 10^{15}\mathrm{g\ s^{-1}}\nonumber \\ &\cdot
\left(\frac{\lambda}{0.1}\right)\left(\frac{M}{10\ \mathrm{M_{\sun}}}\right)^2\left(\frac{\rho}{10^3\ \mathrm{cm^{-3}}\ m_p}\right)\left[\frac{\upsilon^2+c_s^2}{(10\ \mathrm{km\ s^{-1}})^2}\right]^{-3/2} \ ,
\label{eq:accretion_rate}
\end{align}
where $G$ is the gravitational constant, $M$ the BH mass, $\rho$ the
gas mass density, $m_p$ the proton mass, $\upsilon$ the speed of the
BH relative to the interstellar gas, and $c_s$ the effective sound
speed taken from Table \ref{tab:ISMtable}.

To obtain the luminosity we apply the treatment of \citet{MT05}, which
takes into account the transition from the standard disc to the RIAF
(radiatively-inefficient accretion flow) mode in low accretion rate
regime. At high accretion rates the BH accretion is
described with the standard disc model, where the radiation efficiency
$\eta \equiv L / (\dot M c^2)$ 
is constant and the luminosity is proportional to the
accretion rate. However when the accretion rate drops below a
threshold, the disc would switch to the RIAF phase and $\eta$ becomes
proportional to the accretion rate, making the luminosity
proportional to the square of the accretion rate
\citep{NY95,KFM08}. The threshold is expected to be around
1/10 of the Eddington accretion rate, and hence 
\begin{align}
\dot{M}_{\rm th} = \epsilon_{\rm th} \dot{M}_{\mathrm{Edd}}
=1.4\times 10^{18}\mathrm{g\ s^{-1}} 
\left(\frac{M}{10\ \mathrm{M_{\sun}}}\right) 
\left( \frac{\epsilon_{\rm th}}{0.1} \right)
\left(\frac{\eta_{\mathrm{std}}}{0.1}\right)^{-1}, 
\label{eq:accretion_threshold}
\end{align}
where $\dot M_{\rm Edd} \equiv L_{\rm Edd} / (\eta_{\rm std} c^2)$ is the Eddington
accretion rate corresponding to the Eddington luminosity, and
$\eta_{\rm std}$ is the radiation efficiency in the standard disc
regime. Then requiring that $\eta$ changes continuously around the threshold, we model
\begin{align}
\eta=
\begin{cases}
\eta_{\rm std} (\dot{M}/\dot{M}_{\rm th})
&(\mathrm{when}\  \dot{M}< \dot{M}_{\rm th}) \\
\eta_{\rm std} &(\mathrm{when}\ 
  \dot{M}_{\mathrm{th}} < \dot{M} < 2 \dot{M}_{\rm Edd}).
\end{cases}
\label{eq:eta}
\end{align}

We adopt the standard values of $\epsilon_{\rm th} = 0.1$ and
$\eta_{\rm std} = 0.1$ in all of our calculations in this work,
and uncertainties about these parameters are discussed in 
in Section \ref{sec:discussion}.
The bolometric luminosity is then calculated as
\begin{align}
L = & \eta \dot{M}c^2\nonumber \\
= & 3.4 \times 10^{37} \mathrm{erg\ s^{-1}}\nonumber \\
\cdot &
\eta \lambda
\left(\frac{M}{10\ \mathrm{M_{\sun}}}\right)^2\left(\frac{\rho}{10^3\ \mathrm{cm^{-3}}\ m_p}\right)\left[\frac{\upsilon^2+c_s^2}{(10\ \mathrm{km\ s^{-1}})^2}\right]^{-3/2} \ , 
\end{align}
which becomes 
\begin{align}
L = & 9.0 \times 10^{32} \mathrm{erg\ s^{-1}}\nonumber\\
\cdot &
\left(\frac{\lambda}{0.1}\right)^2
\left(\frac{M}{10\ \mathrm{M_{\sun}}}\right)^3\left(\frac{\rho}{10^3\ \mathrm{cm^{-3}}\ m_p}\right)^2\left[\frac{\upsilon^2+c_s^2}{(10\ \mathrm{km\ s^{-1}})^2}\right]^{-3} 
\label{eq:Xray_luminosity_RIAF}
\end{align}
in the RIAF regime and
\begin{align}
L = & 3.4 \times 10^{35} \mathrm{erg\ s^{-1}}\nonumber\\
\cdot & \left(\frac{\lambda}{0.1}\right)
\left(\frac{M}{10\ \mathrm{M_{\sun}}}\right)^2\left(\frac{\rho}{10^3\ \mathrm{cm^{-3}}\ m_p}\right)\left[\frac{\upsilon^2+c_s^2}{(10\ \mathrm{km\ s^{-1}})^2}\right]^{-3/2} 
\label{eq:Xray_luminosity_standard_disk}
\end{align}
in the standard disc regime. 

When the accretion rate largely exceeds the Eddington limit, the
accretion flow would be described by the slim disc rather than the
standard disc. In this regime we adopt the formula by
\citet{Watarai00}:
\begin{align}
L = 2 L_{\rm Edd} 
\left[1+\ln\left(\frac{\dot{M}/\dot{M}_{\rm Edd}}{2}\right)\right] 
\label{eq:Xray_luminosity_Watarai}
\end{align}
for $\dot{M} > 2 \dot{M}_{\rm Edd}$, which is smoothly 
connected to the standard disc regime when
$\eta_{\rm std} = 0.1$ is assumed.

\subsection{IBH Spectrum}
\label{ssec:IBH_spectrum}

Although the spectrum of IBHs is essentially unknown,
  past studies \citep[e.g.,][]{Agol02,Fender13} assumed that their
  characteristics are similar to those observed from BH binaries.
  The spectrum of BH binaries is divided into two categories depending
  on the accretion rate: the soft state and the hard state
  \citep{KFM08}. BH binaries are considered to show the soft state
  spectrum in the standard disk regime (i.e. when the accretion rate
  is high), and the radiation spectrum is dominated in X-rays by a multi-temperature black body radiation from an optically-thin accretion disk. The hard state is typical for low-accreting BHs in the RIAF regime, and
  the radiation is described as a power law with a photon index of
  $\zeta = 1.4$--$2.1$ \citep{RM06}, 
where the differential photon spectrum is $dF_{\rm ph}/d\epsilon_{\rm ph}
\propto \epsilon_{\rm ph}^{-\zeta}$.  

Due to the low density of the interstellar matter, we
  expect that a majority of IBHs are in the RIAF
  regime. Therefore, following \citet{Fender13}, we assume that the
  IBH spectrum is a power-law with $\zeta = 1.6$.  We will see in
  Section \ref{ssec:Flux_and_Properties} that most of the detectable
  IBHs are indeed in the RIAF regime. We assess quantitatively in
  Section \ref{sec:discussion} how our results would change by varying
  the photon index. We also assume that the bolometric luminosity $L$
  is dominantly radiated in the X-ray band of 0.1--100 keV, and then
  the fraction $f_{\rm band}$ of the luminosity in the observed band
  is $f_{\rm band} = 0.31$ for the {\it NuSTAR} and {\it FORCE} bands
  ($10$ -- $40$ keV), and $f_{\rm band} = 0.17$ for the {\it ROSAT}
  band ($0.1$ -- $2.4$ keV).

\subsection{Absorptions}

Furthermore we introduce two parameters, $f_{\rm MC}$ and $f_{\rm
  MW}$, to take into account the photoelectric absorption of X-rays.
When an IBH is accreting in a molecular cloud, the flux is reduced by
a factor of $f_{\rm MC}$ by absorption within the cloud.  We calculate
$f_{\rm MC}$ assuming a hydrogen column density of $N_H=5\times
10^{21} {\rm \ cm^{-2}}$, which is calculated from the density-size
relation of molecular clouds \citep{Larson81}.  According to this
relation, $N_H$ is not sensitive to the size of molecular clouds,
though there exists a scatter from the mean relation by up to an order of
magnitude. We calculate using the model of \citet{Wilms00} that there is a
significant photoelectric absorption of $f_{\rm MC} \sim 0.3$ in the {\it ROSAT} band, but $f_{\rm MC} \sim 1$ for other satellites.  X-rays
are absorbed also by ISM in the Galaxy along the line of sight to the
observer by a factor of $f_{\rm MW}$.  The value of $f_{\rm MW}$
depends on the location of an IBH, and if it is located in GC, X-rays
in the {\it ROSAT} band are seriously absorbed by $f_{\rm MW} \sim
0.01$ with a large column density of $N_H \sim 6\times 10^{22}{\rm
  cm^{-2}}$ \citep{Baganoff03, Muno09}. On the other hand, the
absorption is negligible in the hard X-ray {\it NuSTAR} and {\it
  FORCE} band. These are taken into account
when our results are compared with observational constraints. 

Finally, we assume that the emission is isotropic, and flux measured
on the Earth is calculated assuming that the Sun is located in the
mid-plane (i.e., $z = 0$) at the Galactocentric distance of $R_0 =
8.3$ kpc.


\section{Results}
\label{sec:results}


\subsection{Distribution of IBHs in the Milky Way}
\label{ssec:distribution_result}

For a given set of model parameters, we generate typically $N_{\rm MC}
= 10^6$--$5\times 10^6$ IBHs in our Galaxy with physical quantities obeying
the distributions described in previous sections by the Monte Carlo
method.  The number of generated IBHs is larger for larger
$\upsilon_{\mathrm{avg}}$ because of its higher probability of escaping
from the Galaxy potential. Then the final estimate of detectable IBHs
will be scaled to match the real number of IBHs in the Galaxy, $N_{\rm
  IBH}$.

\begin{figure*}
\begin{tabular}{cc}
\begin{minipage}{0.5\hsize}
 \centering
\includegraphics[width=1.0\linewidth]{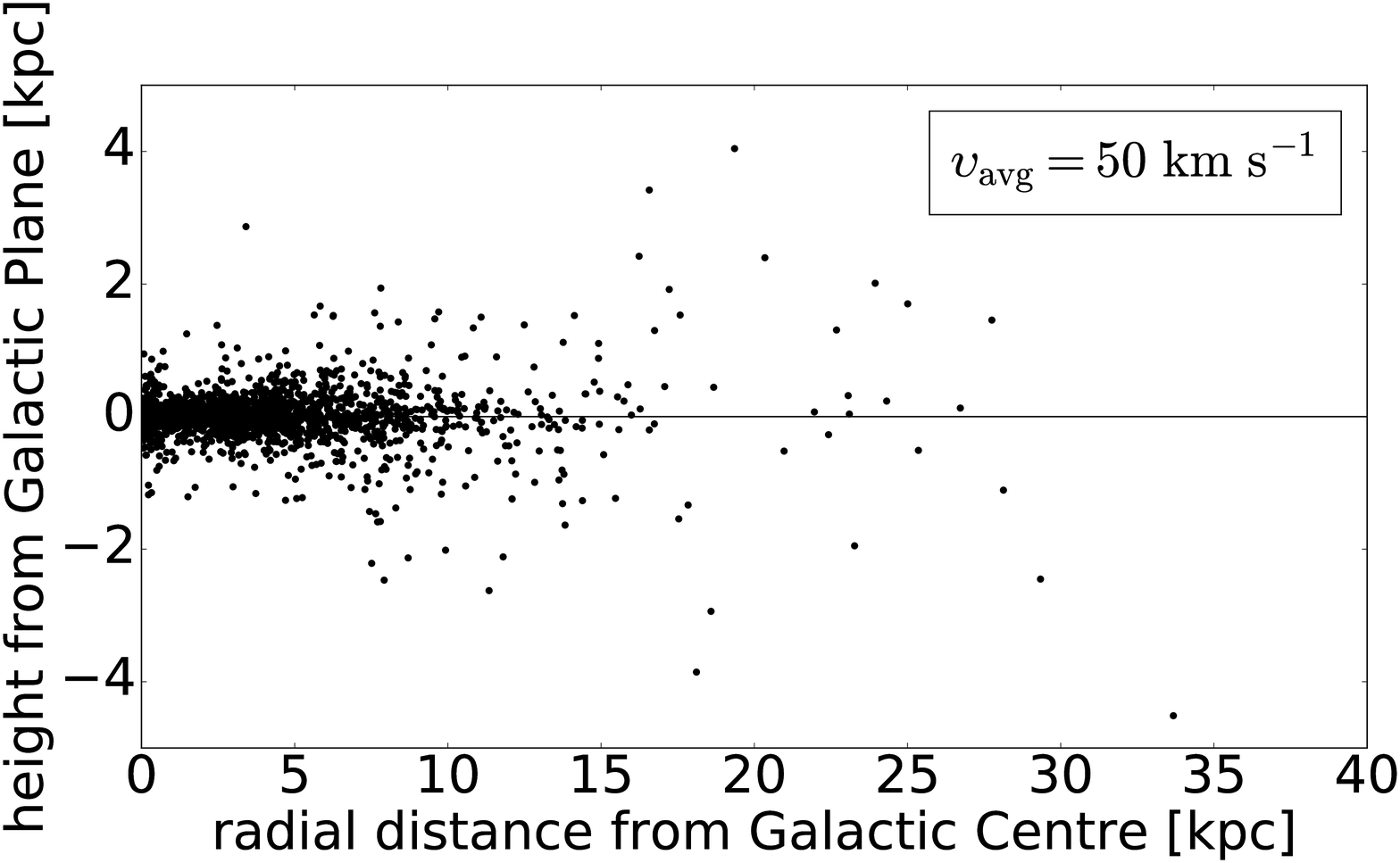}
\end{minipage}
\begin{minipage}{0.5\hsize}
 \centering
\includegraphics[width=1.0\linewidth]{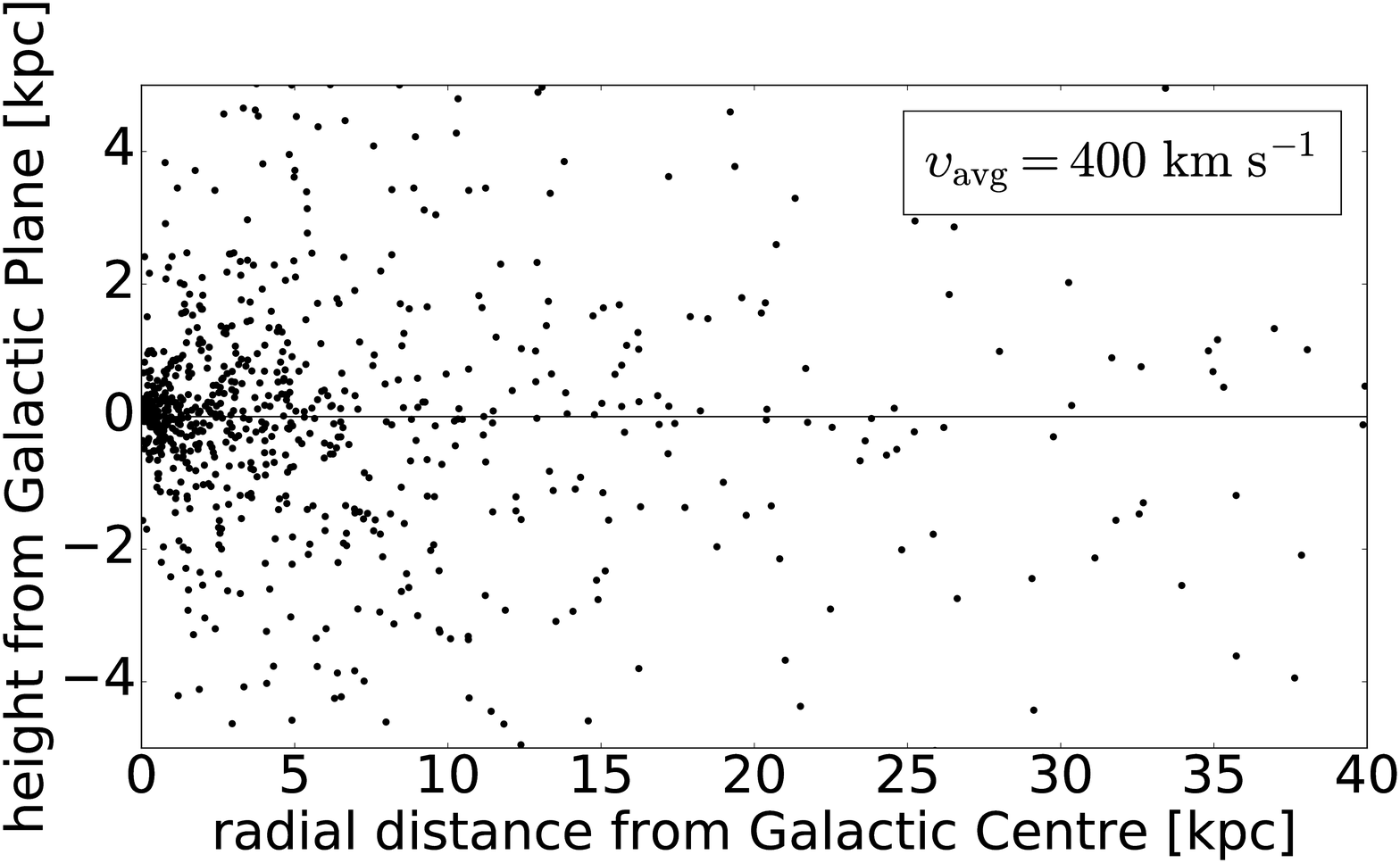}
\end{minipage}
\end{tabular}
\caption{Present IBH distributions for average kick velocity of
  $\upsilon_{\rm avg}$ of $50\ \mathrm{km\ s^{-1}}$ (left) and
  $400\ \mathrm{km\ s^{-1}}$ (right). Only 2000 IBHs 
  randomly chosen from $\sim 10^6$ IBHs actually computed
are shown here.}
\label{fig:IBHdistribution}
\end{figure*}

The present spatial distribution of IBHs in the Galaxy, after orbital
calculations described in Section \ref{ssec:EQM}, is shown in Fig.
\ref{fig:IBHdistribution} for two values of average kick velocity,
$\upsilon_{\mathrm{avg}}=50$ and $400$ $\mathrm{km\ s^{-1}}$.  We find
an obvious trend that IBH distribution becomes more extended from the
GC with increasing $\upsilon_{\rm avg}$.
Note that a portion of IBHs have positive total (kinetic plus
potential) energy and eventually escape the Galaxy.  The fraction is
negligibly small for
$\upsilon_{\mathrm{avg}}=50\ \mathrm{km\ s^{-1}}$, whereas it
increases to 0.01, 3, 17, and 37 per cent for $\upsilon_{\mathrm{avg}}=$
$100, 200, 300$ and $400\ \mathrm{km\ s^{-1}}$, respectively.

\begin{figure}
 \centering
\includegraphics[width=1.0\linewidth]{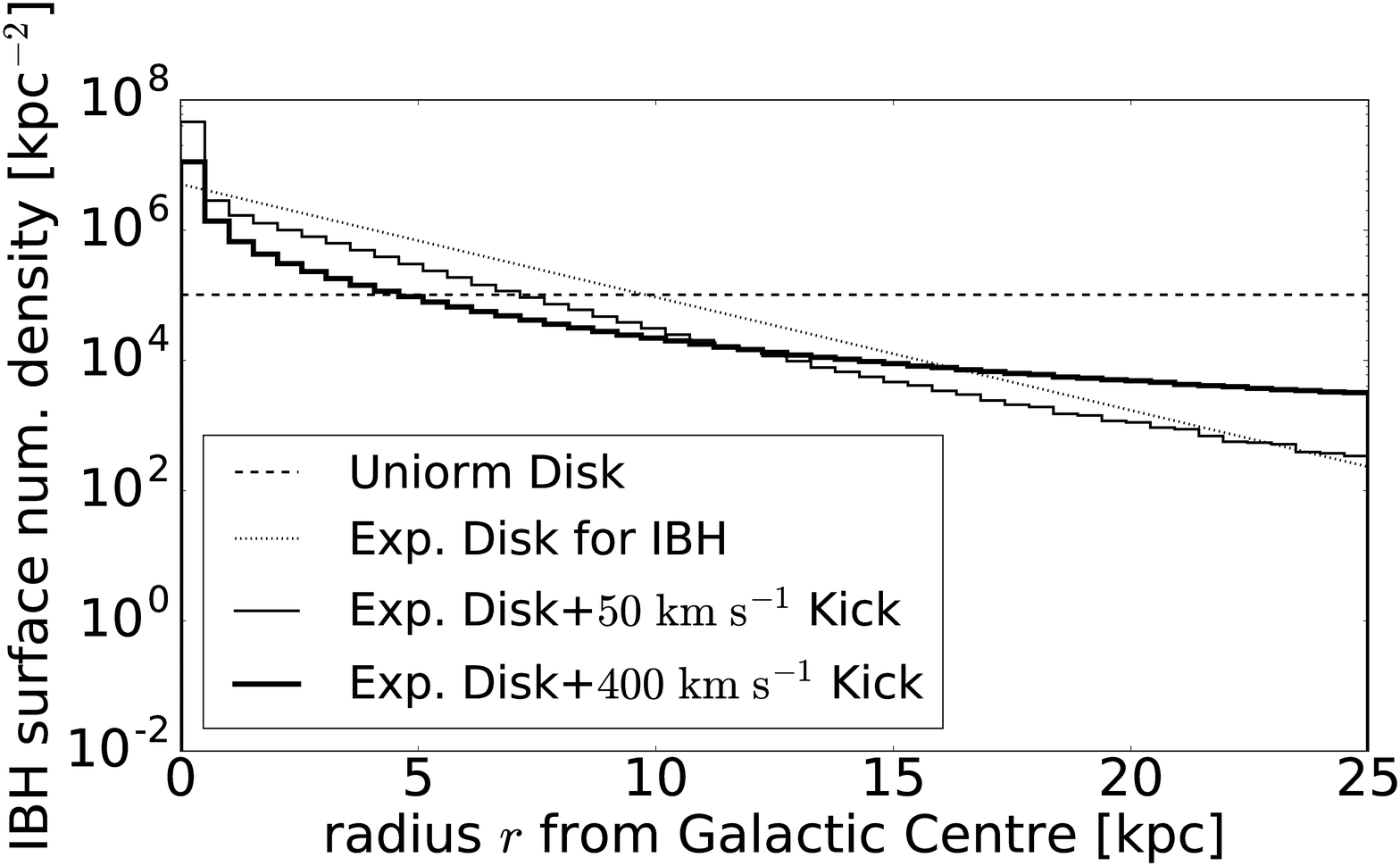}
\caption{The surface number density of IBHs on the Galactic plane.
  Histograms are for present-day IBHs after orbital evolution from
  their birth, with two different values of average kick velocity.
  The total number of IBHs is normalized to $1\times 10^8$. For
  comparison, uniform and exponential (with a scale 2.15 kpc)
  distributions are also shown.}
\label{fig:Comparison_of_IBHdistribution}
\end{figure}

Fig. \ref{fig:Comparison_of_IBHdistribution} shows the surface
number density of IBHs on the Galactic plane, in comparison with the
uniform (i.e. constant surface density) distribution and an
exponential distribution with the scale length of 2.15 kpc, which was
adopted for the initial IBH distribution at their birth.  It can be
seen that the distribution after orbital evolution becomes more
extended than the initial exponential shape, and this effect becomes
stronger with larger kick velocity. It should also be noted that there
is an excess of IBHs near the GC within 1 kpc, which is
due to the contribution from the Galactic bulge component.

\subsection{X-ray Source Counts}
\label{ssec:Flux_and_Properties}

\begin{figure}
\centering
\includegraphics[width=1.0\linewidth]{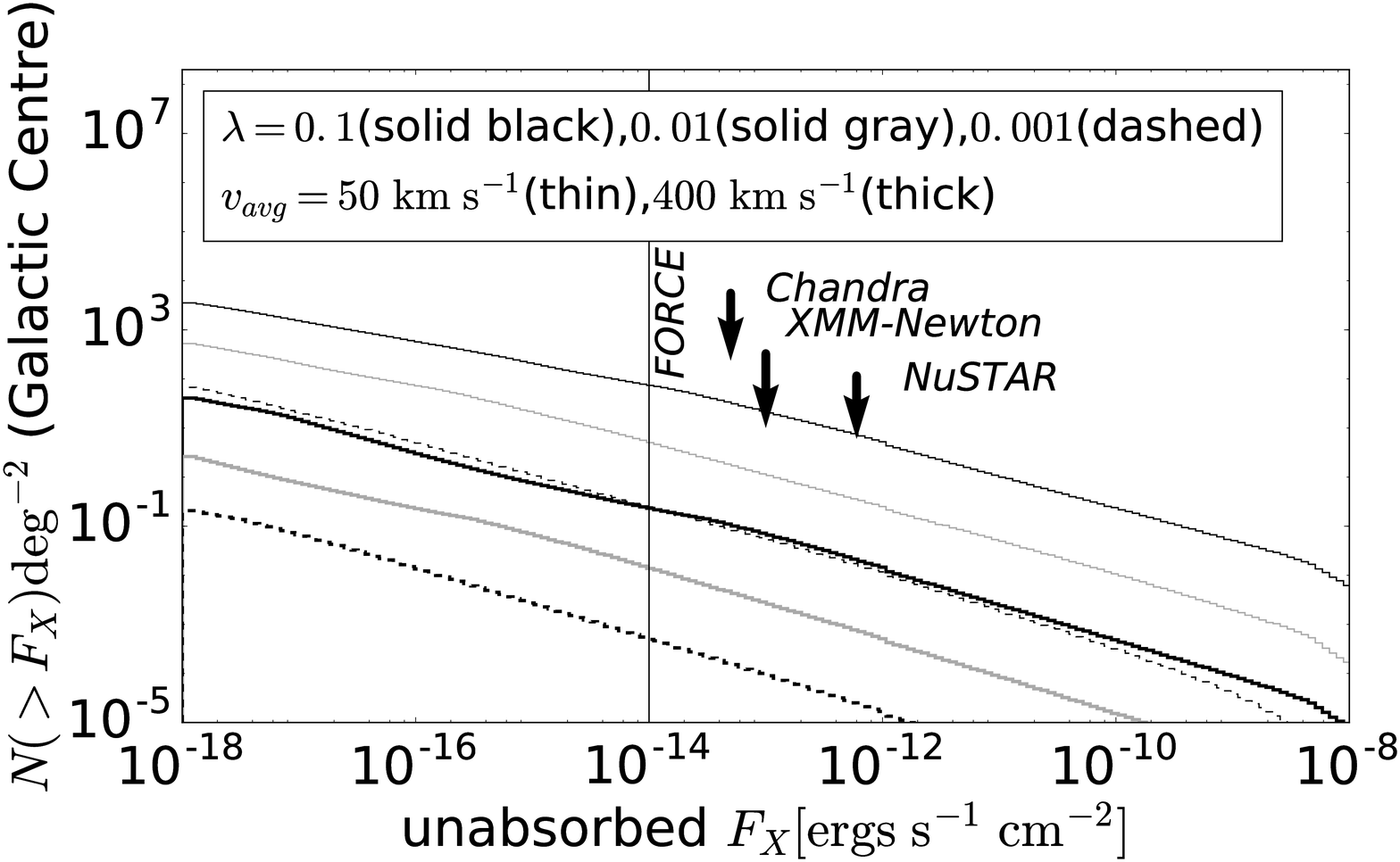} 
\caption{Cumulative X-ray source counts of IBHs with X-ray flux
  greater than $F_X$, inside the 0.5 deg$^2$ region towards the GC
  direction. A hard X-ray energy band of $10$ -- $40$ keV is assumed, and in this energy band the absorptions
  in molecular clouds or ISM are negligible (i.e. $f_{\rm MC}=f_{\rm MW}=1$). Six curves are shown for the
  combination of $\upsilon_{\mathrm{avg}}=50$ or $400
  \ \mathrm{km\ s^{-1}}$, and $\lambda$ = $0.1, 0.01, 0.001$, as
  indicated in the Figure. Conservative upper limits from the
  observations by {\it Chandra} ($0.5$--$8$\ keV), {\it XMM-Newton}
  ($2$--$12$\ keV), and {\it NuSTAR} ($10$--$40$\ keV) are shown, by
  requiring that the number of IBHs should not exceed that of all
  X-ray sources detected toward the GC.  The vertical line shows the
  expected sensitivity of a proposed mission {\it FORCE} in $10$--$40$
  keV energy band. }
\label{fig:histogram_curve}
 \end{figure}

\begin{figure}
\centering
\includegraphics[width=1.0\linewidth]{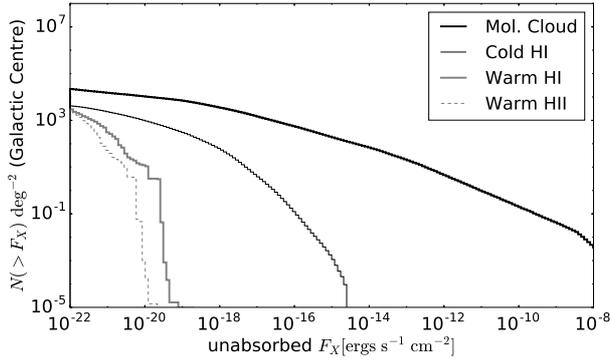} 
\caption{Same as Fig. \ref{fig:histogram_curve}, but broken into
  contributions from each ISM phase.  An average kick velocity of
  $\upsilon_{\rm avg} = 50\ \mathrm{km\ s^{-1}}$ and $\lambda=0.1$ are
  assumed for this figure. It should be noted that the contribution
  from IBHs in the hot H{\sevensize II} ISM phase is too faint to
  appear in this plot.}
\label{fig:histogram_each_phase}
 \end{figure}

\begin{figure}
 \centering
\includegraphics[width=1.0\linewidth]{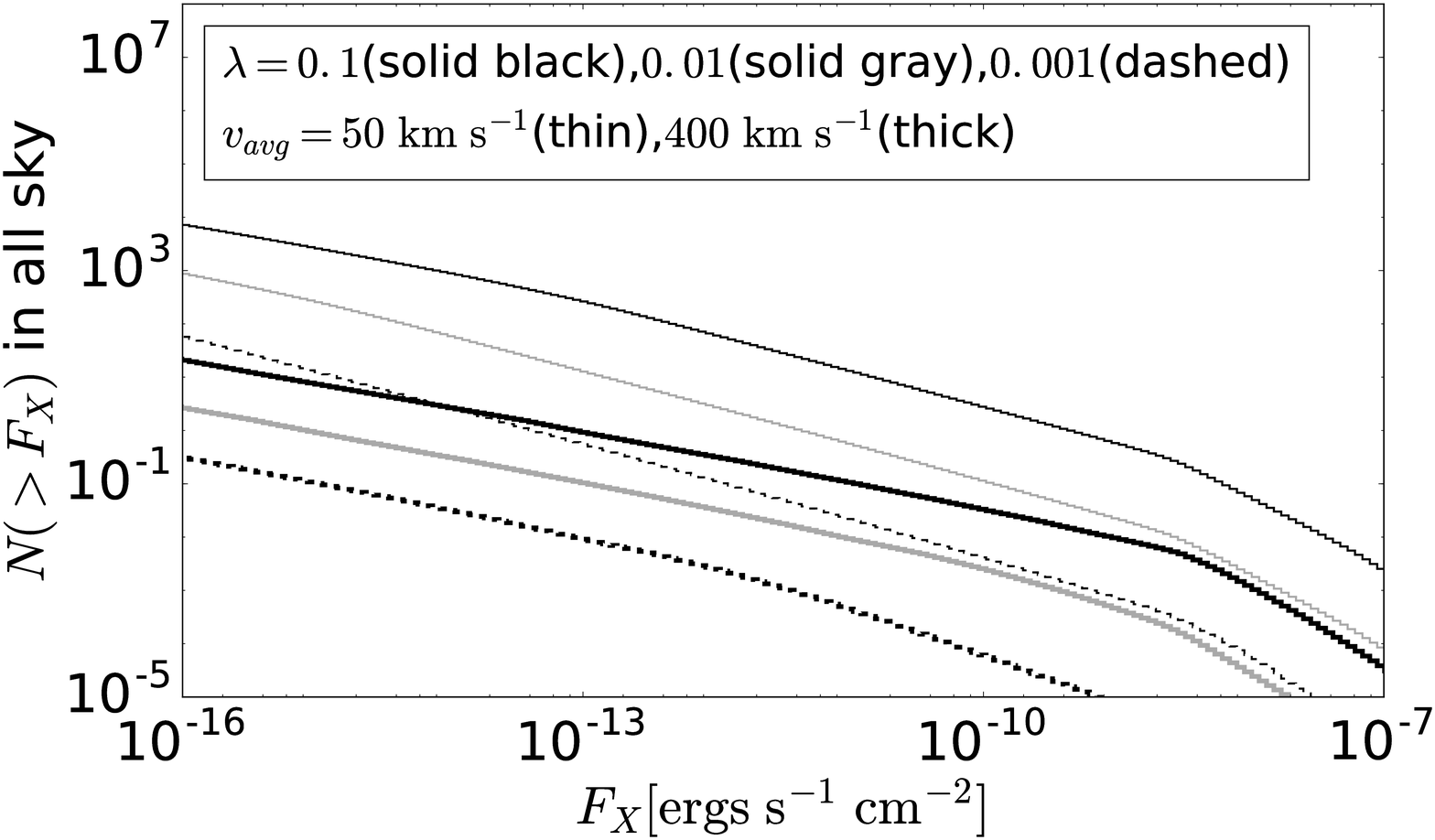} 
\caption{Same as Fig. \ref{fig:histogram_curve}, but for IBHs in all
  sky observed in the {\it ROSAT} 0.1 -- 2.4 keV band. Here we assume the absorption correction as $f_{\rm MC}=0.3$ and $f_{\rm MW}=1$.}
\label{fig:histogram_curve_allsky_minabs}
 \end{figure}

\begin{figure}
 \centering
\includegraphics[width=1.0\linewidth]{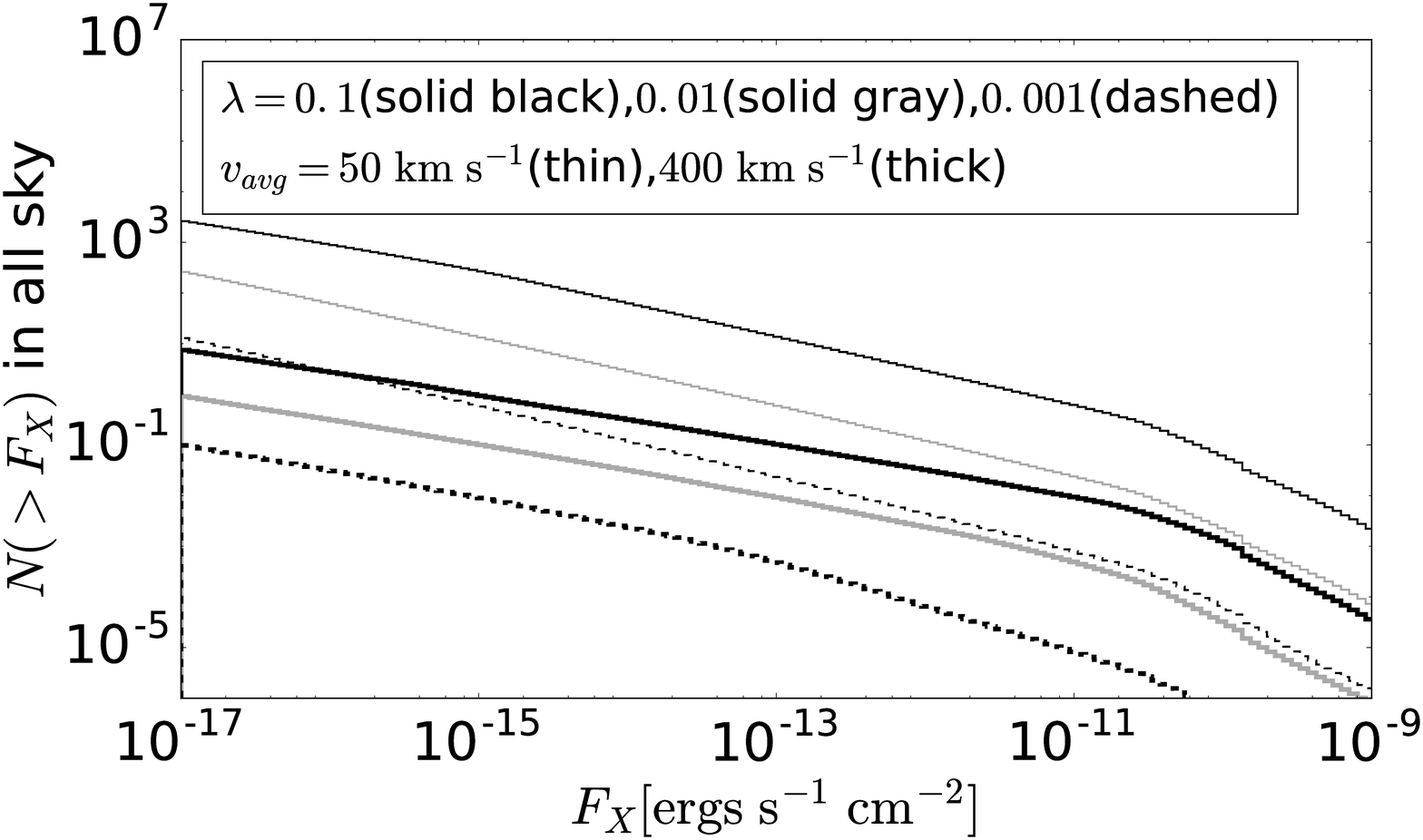} 
\caption{ Same as Fig. \ref{fig:histogram_curve_allsky_minabs}, but $f_{\rm MW}=0.01$ is assumed.}
\label{fig:histogram_curve_allsky_maxabs}
 \end{figure}

Here we present X-ray source counts, i.e., number of IBHs as a
function of X-ray flux. According to the modelling presented in the
previous section, we can assign an X-ray flux for each of the IBHs
whose location and velocity have been calculated by time integration
of their orbits. Figure \ref{fig:histogram_curve}
  shows the cumulative X-ray source counts into the direction of the
  GC, for a few different values of the average kick velocity
  $\upsilon_{\mathrm{avg}}$ and accretion efficiency $\lambda$.  Here
  calculation is for the hard X-ray band where absorptions are
  negligible, i.e., $f_{\rm MC} = f_{\rm MW} = 1$.
  Figure \ref{fig:histogram_each_phase} is the same but shows the
  contribution from each ISM phase in the case of the GC direction.
  Figures \ref{fig:histogram_curve_allsky_minabs} and
  \ref{fig:histogram_curve_allsky_maxabs} are the same as
  Figure \ref{fig:histogram_curve} but for all sky. To compare with the
  {\it ROSAT} result, we assumed $f_{\rm MC} = 0.3$ for the absorption in
  molecular clouds in the {\it ROSAT} band.  The absorption in ISM depends
  on IBH locations, and here we show two extreme cases: no absorption
  ($f_{\rm MW}=1$) and absorption towards GC ($f_{\rm MW}=10^{-2}$) in
  Figures \ref{fig:histogram_curve_allsky_minabs} and
  \ref{fig:histogram_curve_allsky_maxabs}, respectively.

It should be noted that the number of IBHs generated by Monte Carlo
($N_{\rm MC}$) is 1--2 orders of magnitude smaller than the actual
number of IBHs in the Galaxy, $N_{\rm IBH}$, because of the limited
computing time. The results shown in these figures are scaled up to
match the actual number of $N_{\rm IBH}$. One may consider that in
this case our calculation cannot resolve a population of IBHs whose
number is smaller than $N_{\rm IBH}/N_{\rm MC}$ in the Galaxy. However
the results shown in Figures \ref{fig:histogram_curve_allsky_minabs} and \ref{fig:histogram_curve_allsky_maxabs} well extend
to the region of small number ($\ll 1$) in all sky.  This is because
we consider the weight of probability distribution about kick velocity
and gas density in molecular clouds and cold H\,{\sevensize I} ISM. The difference of
$N_{\rm MC}$ and $N_{\rm IBH}$ may also change the distance to the
nearest IBH from the Sun, but it is not important because the nearest
IBHs are not the major component in the detectable IBHs (see Section
\ref{ssec:comparison}, Fig. \ref{fig:distance_from_sun}).

\subsection{Comparison with Observations and Future Detectability}
\label{ssec:comparison}

As a comparison with observations, we first consider the {\it ROSAT}
all-sky survey \citep{ROSAT}. Fig. \ref{fig:histogram_curve_allsky_minabs}
implies that many IBHs may be detected by the {\it ROSAT} sensitivity of
$\sim 1\times 10^{-12} {\rm erg/s/cm^2}$ in all sky with optimistic
model parameters, but it should be noted that the flux plotted in this
figure is not corrected for absorption by intervening ISM, and evidently it is more difficult to detect if maximal absorption is assumed (see Fig. \ref{fig:histogram_curve_allsky_maxabs}).

\begin{figure}
 \centering
\includegraphics[width=1.0\linewidth]{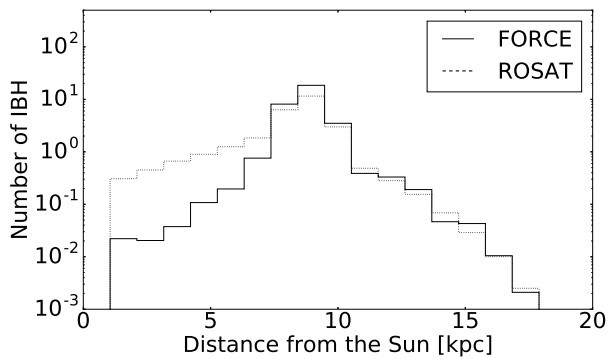} 
\caption{Distribution of distance from the Sun to IBHs detectable by
  the {\it ROSAT} all-sky survey or by the future {\it FORCE} survey
  towards GC of survey area $0.5$ deg$^2$.  Here $\upsilon_{\rm avg}$
  and $\lambda$ are set to $50\ \mathrm{km\ s^{-1}}$ and $0.1$,
  respectively. For {\it ROSAT} the absorption factors of $f_{\rm
    MC}=0.3$ and $f_{\rm MW}=1$ are used like
  Fig. \ref{fig:histogram_curve_allsky_minabs}, but the histogram in reality
  can be significantly modified by adopting location-dependent $f_{\rm
    MW}$.}
\label{fig:distance_from_sun}
\end{figure}

Fig. \ref{fig:distance_from_sun} shows the distribution of distance
from the Sun to IBHs detectable by the {\it ROSAT} all sky survey.  It
can be seen that IBHs towards GC are dominated by those located around
GC.  This is not only because IBHs are concentrated around GC, but
also because the abundance of molecular clouds around GC is high.

This result tells us that considering only the Solar vicinity is not
sufficient for estimating the detectability of IBHs, but it is
necessary to take the entire Galaxy, especially the GC region, into
consideration.  Soft X-ray flux in the {\it ROSAT} band is seriously
reduced by a factor of $f_{\rm MW} \sim 0.01$ due to absorption by ISM
along the sightline to GC, and hence it is difficult to derive a
strong constraint on IBH parameters even if no IBH is included in the
{\it ROSAT} catalog.

For the GC direction, past surveys by {\it Chandra} (flux limit $5
\times 10^{-14} \ \rm erg \ s \ cm^{-2}$ in $0.5$--$8$ keV, survey
area 1.6 deg$^2$), {\it XMM-Newton} ($2.5 \times 10^{-14} \ \rm erg
\ s \ cm^{-2}$ in $2$--$10$ keV, 22.5 deg$^2$), and {\it NuSTAR} ($6
\times 10^{-13} \ \rm erg \ s \ cm^{-2}$ in $10$--$40$ keV, 0.6
deg$^2$) have detected 9017, 2204, and 70 sources, respectively
\citep{Muno09,Warwick11,Hong16}.  Here, the flux limits for {\it
  Chandra} and {\it XMM-Newton} are those corrected for absorption by
ISM, as given in the references. The correction factor is $\sim 3$ for
{\it Chandra}, but it is negligible in the {\it NuSTAR} band.  The
expected number of IBHs with a parameter set of $\lambda = 0.1$ and
$\upsilon_{\rm avg}$ = 50 km/s are 36, $2.4\times 10^{2}$, and 4,
respectively (when $f_{\rm band}=0.3$ for {\it Chandra} and {\it
  XMM-Newton} bands are assumed). Though the number expected for {\it XMM-Newto}n
largely exceeds unity, it is difficult to discriminate IBHs for other populations of X-ray sources. 
Therefore we conservatively set upper
limits that the number of IBHs cannot exceed those of all detected
sources in these surveys.  These upper limits are shown in
Fig. \ref{fig:histogram_curve}, which do not give a strong constraint
on IBH parameters.

\begin{figure}
\centering
\includegraphics[width=1.0\linewidth]{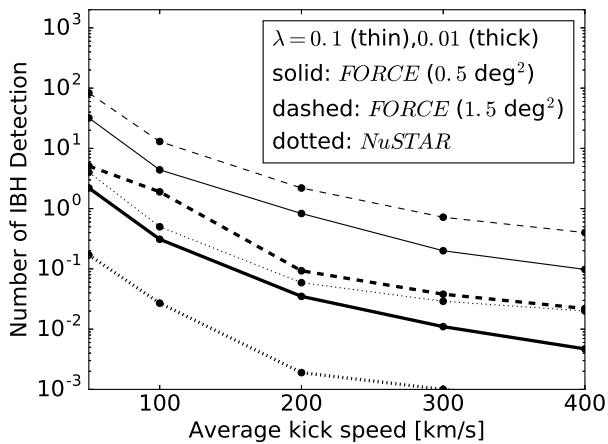} 
\caption{The expected number of IBH detections by a survey by {\it
    FORCE} towards GC, as a function of mean kick velocity
  $\upsilon_{\rm avg}$ for two values of $\lambda$. The sensitivity is
  assumed to be $1\times 10^{-14}\ \mathrm{erg\ s^{-1} cm^{-2}}$ in
  the $10$--$40$ keV, and two different survey area (0.5 and 1.5
  deg$^2$) are considered, as shown by the solid and dashed
  curves. For comparison, the expected number by the past {\it NuSTAR}
  survey \citep{Hong16} is also shown.}
\label{fig:detection_number}
 \end{figure}

As mentioned in the Introduction, hard X-ray band may be useful to
discriminate IBHs from other X-ray populations.  However, the expected
number of IBHs detectable by the past {\it NuSTAR} survey is at most
of order unity. Therefore we consider a survey towards GC by the
proposed mission {\it FORCE}.  A sensitivity limit of about $1\times
10^{-14}\ \mathrm{erg\ s^{-1} cm^{-2}}$ in the $10$--$40$ keV can be
achieved by 100 ksec observation for each field-of-view, by the
improved angular resolution compared with {\it NuSTAR}.  The
sensitivity flux limit is indicated in Fig. \ref{fig:histogram_curve}.
A total survey area of about 1 deg$^2$ is possible with a realistic
telescope time.  We show in Fig. \ref{fig:detection_number} the
expected number of IBH detections by a survey by {\it FORCE} as a
function of $\upsilon_{\rm avg}$ with two values of $\lambda$,
assuming a total survey area of 0.5 and 1.5 deg$^2$.  The expected
number becomes much larger than unity at an optimistic parameter
region of $\lambda = 0.1$ and $\upsilon_{\rm avg} \lesssim 100$ km
s$^{-1}$, and hence no detection of an IBH by this survey can exclude
this parameter region, provided that discrimination of IBHs from other
X-ray source populations is successful.

\subsection{Interpretation of Source Counts}
\label{ssec:interpretation}

The shape of the X-ray source counts in Fig.
\ref{fig:histogram_curve} can be understood as follows.  The shape of
$\lambda = 0.1$ curves is a broken power-law with the break flux of
$\sim 3 \times 10^{-14}\ \mathrm{erg\ s^{-1} cm^{-2}}$. The power
index is $\alpha\sim 0.5$ and $0.7$ at the lower and higher flux
regimes, respectively, where $N(>F_X)\propto F_X^{-\alpha}$.  As seen
in Fig. \ref{fig:histogram_each_phase}, source counts are dominated by
IBHs in molecular clouds. When velocity of an IBH is less than $c_s$,
accretion rate and luminosity are determined by gas density $n$ with a
weak dependence on $\upsilon$ (see eq. \ref{eq:Xray_luminosity_RIAF}).
Since the probability distribution of gas density is a power-law,
$d\xi/dn\propto n^{-2.8}$, the probability of finding a density larger
than $n$ is $P(>n) \propto n^{-1.8}$. In addition, the number of black
holes that are lower than $c_s (\ll v_{\rm avg})$ scales as
$N(<c_s)\propto c_s^3 \propto n^{-1.05}$ due to the assumed
Maxwell-Boltzmann distribution of kick velocity. In this regime the
IBH luminosity is approximately proportional to 
$n^2 c_s^{-6} \propto n^{4.1}$ by
eq. (\ref{eq:Xray_luminosity_RIAF}), and we obtain $\alpha\sim
(1.8+1.05)/4.1 \sim 0.7$, which is consistent with the slope observed
at fluxes above the break.  

The break flux should correspond to that
for the minimum density of molecular clouds, $n_1 =
10^2\ \mathrm{cm^{-3}}$, with $\upsilon\sim c_s= 3.7\ \mathrm{km
  \ s^{-1}}$. Using these parameters, we find $\sim 3 \times
10^{-14}(\lambda/0.1)^2\ \mathrm{erg\ s^{-1} cm^{-2}}$ at GC for the
mean BH mass of $M=7.8\ \mathrm{M_{\sun}}$, which is consistent with
the break flux found in the figure. On the other hand, the sharp drop
of the counts at the bright end of $F_X \sim
10^{-8}\ \mathrm{erg\ s^{-1}\ cm^{-2}}$ common for all curves corresponds to the transition
luminosity, where the RIAF regime switches into the standard disc
regime. The luminosity here is equivalent to $0.1$ times the Eddington
luminosity, which gives $F_X \sim
10^{-8}\ \mathrm{erg\ s^{-1}\ cm^{-2}}$ when an IBH at GC is assumed.

The source counts below the break flux is dominated by IBHs having
velocities larger than $c_s$ and hence fainter flux.  When $\upsilon >
c_s$ but still $\upsilon < \upsilon_\mathrm{avg}$, The number of IBHs
with a velocity lower than $\upsilon$ roughly scales as
$N(<\upsilon)\propto \upsilon^3$. Therefore $\alpha\sim 0.5$ 
should be found in this range, which is also 
consistent with the curves in the figure.

The source counts in all sky (Figures \ref{fig:histogram_curve_allsky_minabs} and \ref{fig:histogram_curve_allsky_maxabs})
show similar breaks and drops to those found in the counts
towards GC. However they appear more smoothed out because distances of
IBHs from the Sun are more widely distributed with less concentration
to GC, as seen in Fig. \ref{fig:distance_from_sun}.


\subsection{Properties of Detectable IBHs}

\begin{figure*}
\centering
 \begin{tabular}{cc}
 \begin{minipage}{0.5\hsize}
 \centering
\includegraphics[width=1.0\linewidth]{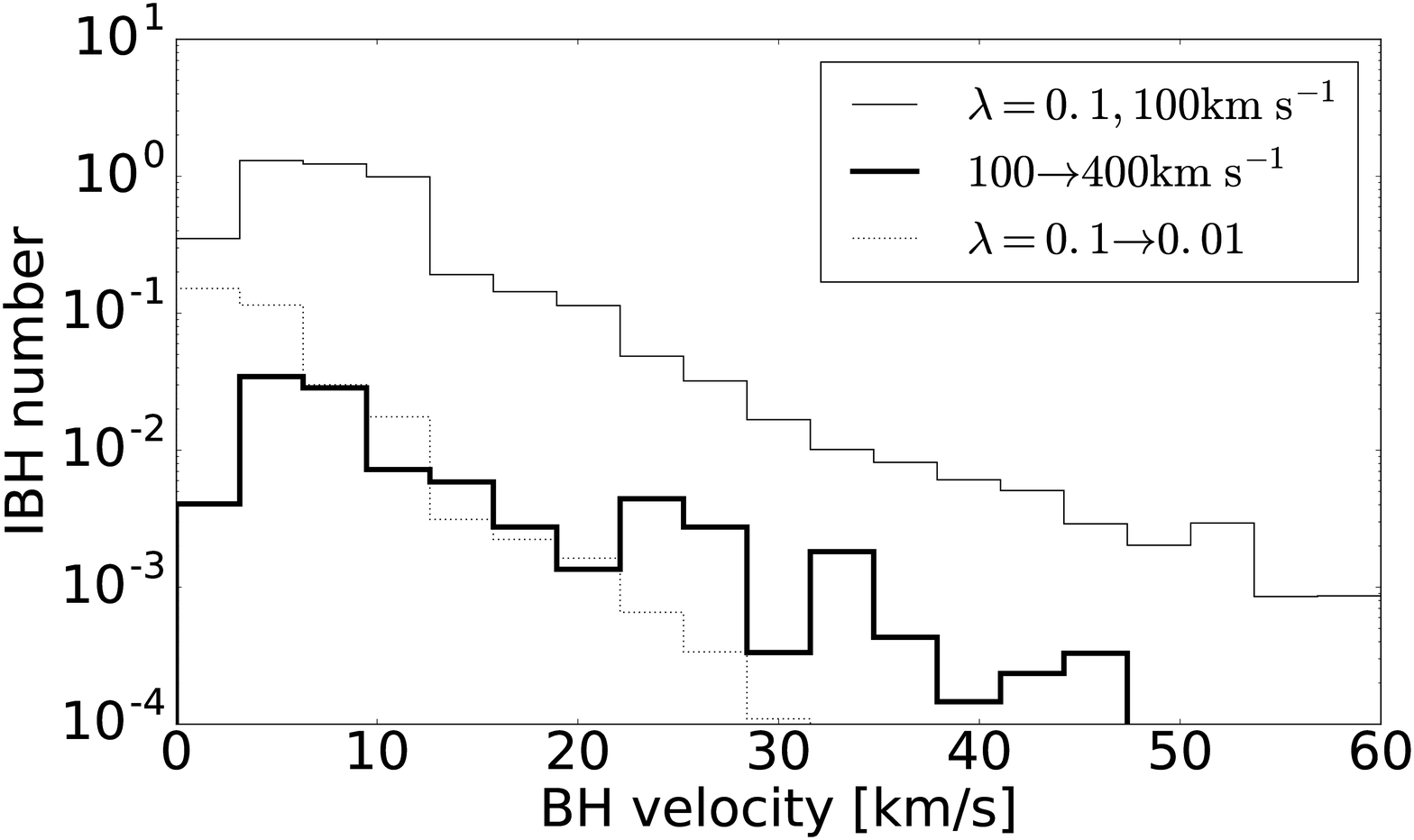} 
 \end{minipage}
\begin{minipage}{0.5\hsize}
 \centering
\includegraphics[width=1.0\linewidth]{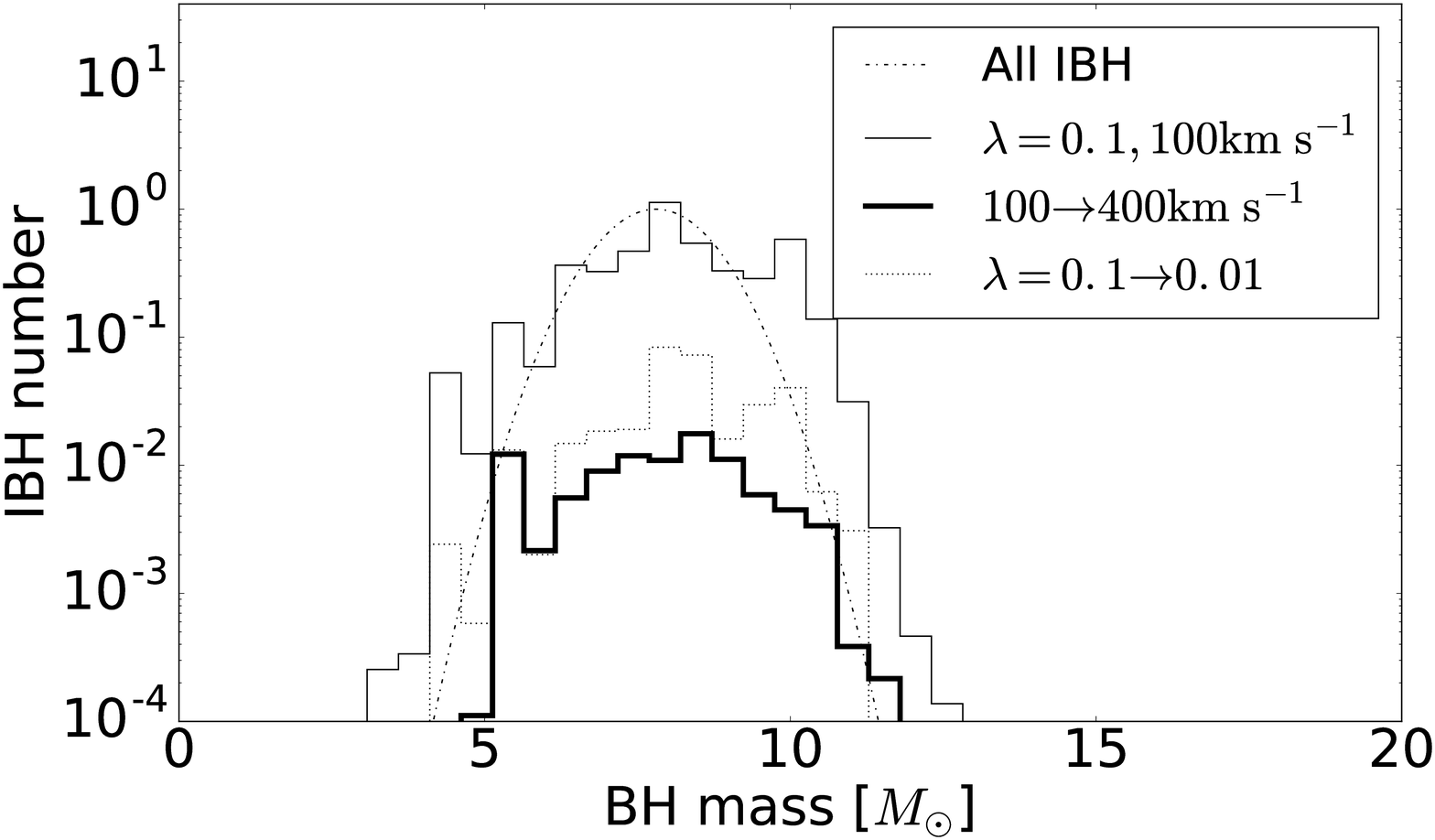}
\end{minipage}
\end{tabular}
\caption{Distributions of IBH velocities $\upsilon$ (with respect to
  the frame of the Galactic rotation) (left) and IBH masses (right),
  for IBHs detectable with {\it FORCE} towards GC with a survey area of $0.5$ deg$^2$. The assumed model
  parameters of $\lambda$ and $\upsilon_{\rm avg}$ are indicated in
  the panels.  The thin dot-dashed curve in the right panel is the
  mass distribution of the entire IBH population \citep{Ozel10}, which
  is scaled for comparison. 
}
\label{fig:physical_properties}
\end{figure*}

Fig. \ref{fig:physical_properties} shows the distribution of IBH
velocities $\upsilon$ (with respect to the frame of the Galactic
rotation) and IBH masses, for IBHs detectable by the future {\it
  FORCE} survey towards GC. As expected from eq. \ref{eq:Xray_luminosity_RIAF}, detectable IBHs are dominated by those
with low velocities. The velocity distribution becomes flatter when
$\upsilon$ becomes close to the effective sound speed of the
molecular gas ($c_s \lesssim 10\ \mathrm{km\ s^{-1}}$).  On the other
hand, the observable IBH mass distribution is almost the same as that
of the entire IBH population, though the peak of the former is
slightly shifted to larger mass. This is because IBH masses are
narrowly distributed, although X-ray luminosity is proportional to the
cube of BH mass in the RIAF regime (eq. \ref{eq:Xray_luminosity_RIAF}).

It is interesting to see the relative contributions to detectable IBHs
from those formed in the Galactic disc or bulge.  This is shown as the
expected number of IBHs detectable by the future {\it FORCE} survey,
for several values of $\upsilon_{\rm avg}$ and assuming $\lambda =
0.1$, in Table \ref{tab:bulge_or_disk_0.1}.  We see that the disc
fraction becomes smaller as $\upsilon_{\rm avg}$ increases, because
IBHs in the disc region are more efficiently expelled by a large kick
velocity from dense gas regions than those in the bulge.  Since the
expected number of bulge IBHs is always much less than unity, IBHs
detectable by {\it FORCE} would be mostly of the disc origin, though
they are located around GC.

\begin{table}
\centering
\begin{tabular}{|c|c|c|c|c|c|}
\hline
& \multicolumn{5}{|c|}{$\upsilon_{\mathrm{avg}}\  [\mathrm{km\ s^{-1}}]$} \\ 
\cline{2-6}
 & $50$ & $100$ & $200$ & $300$ & $400$ \\ \hline 
bulge & $0.018$ & $1.4\times 10^{-3}$ & $0.067$ & $3.7\times 10^{-3}$ & $9.1\times 10^{-3}$ \\ \hline
disc & $32$ & $4.5$ & $0.79$ & $0.19$ & $0.086$ \\ \hline
\end{tabular}
\caption{Expected number of IBHs detectable by the future {\it FORCE}
survey, showing separately those born in the Galatctic disc and bulge. 
For all calculations $\lambda=0.1$ is assumed.
}
\label{tab:bulge_or_disk_0.1}
\end{table}


\section{Discussion}
\label{sec:discussion}

In this section we mention some caveats and important notes of our work.

\subsection{Parameters for Radiative Efficiency of Accretion Flow}
The parameter $\epsilon_{\rm th}$ in Section \ref{ssec:lum_flux} is
  sensitive to the viscosity parameter $\alpha$. \cite{NY95} have shown
  that $\epsilon_{\rm th}$ can become approximately $10^{-3}$ --
  $10^{-1}$ with $\alpha=0.03$--$0.3$.
  When $\epsilon_{\rm th}$ smaller than our value $10^{-1}$ is assumed, the radiation efficiencies of IBHs whose accretion rates are smaller than the threshold would be larger. This increases the luminosities of IBHs in the RIAF regime, 
thus significantly increasing the number of detections. For example, by changing $\epsilon_{\rm th}$ to
  $10^{-3}$ we find that the number of IBHs detectable by {\it FORCE}
is increased to $2.8\times 10^{2}$ by a factor of about ten,
  assuming $\upsilon_{\rm avg} = 50$ km s$^{-1}$, $\lambda = 0.1$,
  and the survey area $0.5$ deg$^2$.
The parameter $\eta_{\rm std}$ is likely less uncertain, 
which is known to be 0.06 for a 
Schwarzschild black hole and 0.4 for a maximally rotating Kerr black
holes \citep[e.g.,][]{Thorne74}.

\subsection{IBH Spectrum}
Considering a different X-ray spectrum of IBHs will change
  $f_{\rm band}$ of each telescope and consequently the flux observed
  in each band. We have assumed a power-law spectrum because
  detectable IBHs are almost always in the RIAF regime, and if we
  consider the photon index to be $\zeta = $1.4 (2.1) and maintain all
  the other assumptions, $f_{\rm band}$ of {\it ROSAT} and {\it FORCE}
  changes to 0.092 (0.55) and 0.33 (0.16), respectively, from the
  values of $f_{\rm band} = 0.17$ and $0.31$ for $\zeta = 1.6$.  From
  the discussions in Section \ref{ssec:interpretation}, the number of
  detections by {\it ROSAT} and {\it FORCE} changes by the flux limit
  to the power of at most 0.7. Then the change of detectable
IBH number is at most a factor of two.

\subsection{Identification of IBHs}
If an IBH candidate is found in hard X-ray surveys
  like {\it FORCE}, it is then necessary to distinguish from other
  sources emitting X-rays, such as cataclysmic variables (CVs) or
  X-ray binaries. Soft and/or thermal sources, such as CVs, may be
  discriminated by selecting hard power-law spectrum sources as
  expected for RIAF. Discrimination from X-ray binaries may be
  possible by looking for a binary companion using infrared
  telescopes \citep[e.g.,][]{Matsumoto17}. Our calculation predicts that IBHs will be found
  preferentially in dense molecular clouds, and IBH candidates would
  be particularly strong if they are embedded in molecular clouds
  around GC found by radio observations.  Background AGNs may be
  removed by checking variability time scales and long-term proper
  motions \citep{Fender13}.  The purpose of this work is to predict the number of IBHs
  that are bright enough to be detected, and detailed observational
  strategies to identify them are beyond the scope of this paper.


\section{Conclusions}
\label{sec:conclusions}

In this work we estimated the number and X-ray luminosity of IBHs
accreting from ISM or molecular cloud gas, and investigate
detectability by past and future surveys, by taking into account the
realistic structure of our Galaxy.  The orbit of each IBH is
calculated by integrating the equation of motion in the Galactic
potential, and luminosity is calculated considering various phases of
ISM and molecular clouds. An important result is that most of the
detectable IBHs reside near the GC, not only for surveys targeted to
GC but also for the {\it ROSAT} survey in all sky.  This demonstrates
the importance of considering the entire Galactic profile in a search
for IBHs.

The detectable number of IBHs was calculated with two main model
parameters: the average of kick velocity $\upsilon_{\mathrm{avg}}$ and
the ratio of actual accretion to the Bondi accretion rate $\lambda$.
We found that a few tens of IBHs would be detected by {\it ROSAT} with
an optimistic parameter set of $\upsilon_{\mathrm{avg}}$ = 50 km
s$^{-1}$ and $\lambda = 0.1$, if we ignore absorption. However, most
of such IBHs are in GC and soft X-rays should be severely absorbed in
ISM, and hence non-detection of IBHs by {\it ROSAT} does not give a strong
constraint.  The expected number for the survey by {\it XMM-Newton}
towards GC is a few hundred for the same parameter set, and IBHs may
have been detected in the {\it XMM-Newton} sources, though
discrimination from other X-ray source populations may be
difficult. The hard X-ray band may have an advantage about
discrimination, but the expected number for the survey performed by
{\it NuSTAR} is at most order unity ($\sim$4).  The future {\it FORCE}
survey towards GC in hard X-ray may detect 30--100 with the same 
parameter set, although this depends on the actual survey
parameters.  It should be noted that $\lambda$ can be smaller than 0.1
depending on the physics of accretion, and in such a case IBH
detection may be difficult even in the foreseeable future.

\section*{Acknowledgements}
The authors express their deepest thanks to Masayoshi Nobukawa and
Koji Mori for information on the {\it FORCE} satellite and many other
valuable comments. We thank the anonoymous referee for suggestions which greatly improved the manuscript of this work. 
We also thank the faculty members at the University of Tokyo, notably Noriyuki Matsunaga, Toshikazu Shigeyama, Hideyuki
Umeda, Kazuhiro Shimasaku, and Kazumi Kashiyama, for fruitful comments
and discussions.  NK is supported by the Hakubi project at Kyoto University. 
TT was supported by JSPS KAKENHI Grant Numbers JP15K05018 and JP17H06362.

\bibliographystyle{mnras} \bibliography{IBH}

\bsp	
\label{lastpage}
\end{document}